\documentclass{aa}
\usepackage{natbib}
\usepackage{graphicx}
\usepackage{amssymb}
\usepackage{amsmath}

\begin{document}

\newcommand{\water}{H$_2$O }
\newcommand{\aori}{$\alpha$~Orionis }
\newcommand{\mucep}{$\mu$~Cephei }

\title{Amorphous alumina in the extended atmosphere of \aori}

\author{
T.~Verhoelst \inst{1,2,\star}
\and L.~Decin \inst{1} \fnmsep \thanks{Postdoctoral Fellows of
the Fund for Scientific Research, Flanders} 
\and R.~Van Malderen \inst{1}
\and S.~Hony \inst{1}
\and J.~Cami \inst{3}
\and K.~Eriksson \inst{4} 
\and G.~Perrin \inst{2}
\and \\ P.~Deroo \inst{1}
\and  B.~Vandenbussche \inst{1} 
\and L.B.F.M.~Waters \inst{1,5}
}

\institute{
Instituut voor Sterrenkunde, K.U. Leuven, Celestijnenlaan 200B, B-3001
Leuven, Belgium
\and Observatoire de Paris-Meudon, LESIA, 5 place Jules Janssen, 92195
Meudon, France
\and NASA Ames Research Center, MS 245-6, Moffett Field, CA 94035
\and Institute for Astronomy and Space Physics, Box 515, 75120
Uppsala, Sweden 
\and Astronomical Institute ``Anton Pannekoek'', University of
Amsterdam, Kruislaan 403, 1098 SJ Amsterdam, The Netherlands  
}

\date{Received 3 May 2005 / Accepted 10 October 2005}

\abstract{
In this paper we study the extended atmosphere of the late-type
supergiant $\alpha$~Orionis. Infrared spectroscopy of red supergiants
reveals strong molecular bands, some of which do not originate in the
photosphere but in a cooler layer of molecular material above
it. Lately, these layers have been spatially resolved by near and
mid-IR interferometry. In this paper, we try to reconcile the IR
interferometric and ISO-SWS spectroscopic results on \aori with a
thorough modelling of the photosphere, molecular layer(s) and dust
shell. From the ISO and near-IR interferometric observations, we find
that \aori has only a very low density water layer close above the
photosphere. However, mid-IR interferometric observations and a
narrow-slit N-band spectrum suggest much larger extra-photospheric
opacity close to the photosphere at those wavelengths, even when
taking into account the detached dust shell. We argue that this cannot
be due to the water layer, and that another source of mid-IR opacity
must be present. We show that this opacity source is probably neither
molecular nor chromospheric. Rather, we present amorphous alumina
(Al$_2$O$_3$) as the best candidate and discuss this hypothesis in the
framework of dust-condensation scenarios.

}

\titlerunning{Amorphous alumina in the extended atmosphere of \aori}
\authorrunning{Verhoelst et al.}
\maketitle


\section{Introduction}
\label{sec:introd}

It is now becoming clear that late-type supergiant stars, like their
less massive AGB counterparts, are also embedded in a circumstellar
environment (CSE) of molecular layers, gaseous outflow and dust
\citep[e.g. ][]{Richards1998,Tsuji2000a,Mucep2005}.  What sets them
apart, however, are the low amplitude pulsations, relatively high
effective temperatures and sometimes the presence of a
chromosphere. It is therefore questionable whether the mechanism
driving their mass-loss is similar to that of AGB stars, i.e. a
complex interplay between pulsations, shock waves, molecular opacity
and dust condensation \citep[][ and references
therein]{Fleischer1995,Hoefner2003}.

\aori and \mucep play an important role in this area of research,
since it was in these stars that molecular layers (containing water)
around late-type supergiants were detected for the first time:
\cite{Tsuji1978} found \mucep to be 0.5~mag brighter between 5 and
8~$\mu$m due to emission by hot \water in the circumstellar
environment.  Later, \cite{Tsuji2000b} also found emission lines of
water in the ISO-SWS spectrum of \mucep and determined temperature,
location and column density of this so-called {\sl molsphere}. In a
re-analysis of old Stratoscope II data of $\alpha$~Orionis,
\cite{Tsuji2000a} identified unexpected absorption lines as due to
non-photospheric water.

Their large apparent size makes both supergiants good candidates for
optical/IR interferometry. In recent years, the presence of these
molecular layers was confirmed with optical/IR interferometry,
allowing the determination of their size, temperature and optical
depth at the wavelengths studied \citep{RSG2004,Mucep2005}.

Generally, modelling attempts of the molecular extended atmospheres of
these supergiants have targeted either spectroscopic data or
interferometric data, but not both simultaneously. The one exception
is a study of \aori by \cite{Ohnaka2004}, in which mid-IR
interferometric and spectroscopic data are modelled using a
blackbody+molecular layer approximation.

In this paper we present a model for $\alpha$~Orionis, which fits
near- to mid-IR interferometric and spectroscopic data
simultaneously. Other crucial difference with previous works is that
we take the photospheric molecular bands into account and cover a
larger wavelength regime. In section~\ref{sec:observ} we present the
IR data collected from the literature and the supplementary data used
in our analysis. Sect.~\ref{sec:modelling} describes the models for
photosphere, molecular layers and dust shell used to interpret the
data. In Sect.~\ref{sec:comparison_aori} we compare the models with
data for \aori and discuss the results. Finally, in
Sect.~\ref{sec:conclusions}, we conclude and look ahead.


\section{The observations}
\label{sec:observ}

\subsection{Spectroscopy}
\label{sec:obs:iso}

 \aori was observed with the ISO-SWS (Infrared Space Observatory,
1995--1998, Short Wavelength Spectrometer, $2.38-45.2~\mu$m at a
spectral resolving power of up to R~$\sim 3800$) on MJD$= 50729$ with the
AOT01 template at speed 4. AOT01 refers to a single full-wavelength
up-down scan for each aperture with four possible scan speeds at
degraded resolution \citep{Leech2002}.  Speed~4 is the slowest
scanning velocity, yielding a spectral resolution of $800-1500$.

In total, the SWS uses 4 detector arrays associated with the SW (Short
Wavelength) and LW (Long Wavelength) gratings (SW: $2.38 - 12.0\,\mu$m
and LW: $12.0 - 45.2\,\mu$m), with 12 elements each. Every 12-element
array observes one wavelength-band, with band 1 ranging from $2.38 -
4.08$~$\mu$m, band 2 from $4.08 - 12.0\,\mu$m, band 3 from $12.0 -
29.0\,\mu$m and band 4 from $29.0 - 45.2\,\mu$m (a definition of all
sub bands is given in Table~\ref{table:factors}). The flux is measured
by integrating the photocurrent produced in the photoconductors on an
integrating capacitance on which the light of a certain wavelength is
collected.  The voltage over the capacitance is read out
non-destructively at 24Hz during an integration interval. An
integration interval typically lasts for one or two seconds. At the
end of the integration interval the capacitor is discharged to start a
new integration. The slope of the non-destructive readouts of the
voltage over the integrating capacitance is a measure of the flux
falling onto the detector.

The data reduction procedure we follow uses the IA (Interactive
Analysis) tools also used in the latest version of the pipeline
(OLP10.1, OLP $=$ Off-line Processing), complemented with a manual
removal of glitches, bad detector data points and scan jumps.  For a
detailed description of these procedures, we refer to \cite{Cami2002}
and \cite{VanMalderen2004}. Furthermore, we check the correction for
memory effects \citep{Kester2001} by comparing both up and down scans
with the final spectrum. Joining of the sub bands was done by means of
the overlap between the different bands, and starting from band 1d
which is believed to have the best absolute flux calibration. For a
detailed description of the need for this procedure (problematic dark
current subtraction, pointing errors, etc), we refer to
\cite{VanMalderen2004}.

Below, we discuss the peculiarities of the data reduction.

\paragraph{Presaturation}
$\alpha$ Orionis  brightness (F$_{\nu}\sim 20000$~Jy at 2.5~$\mu$m)
proves to be especially challenging, mainly because the dynamic range
of the detectors was optimised for sources of a few to a few thousand
janksy. While there is no sign of actual saturation of the detectors,
i.e. there is no clear cut-off value, the integration ramp of some of
the detectors shows a strong non-linear behaviour when the voltage
across the integrating capacity reaches very high values. This
flattening of the integration-ramp toward the end of the integration
time causes an underestimation of the slope and therefore of the flux
at that wavelength. A similar phenomenon has been seen by the
SWS-Instrument-Dedicated-Team (SIDT) in a post-Helium observation
(after the satellite had run out of coolant), and is called
``presaturation''.  In the pipeline-reduced data, this presaturation
causes an ``absorption feature'' at 3\,$\mu$m which is not real. 

The solution to this problem of presaturation, presented by the SIDT,
is straightforward: only use that part of the integration-ramp which
behaves linearly. In concreto this comes down to using only the first
half or equivalently the first 24 voltage measurements minus those
rejected for other reasons\footnote{In the case of this AOT01 speed 4
observation, the integration time is 2 seconds and the voltage is read
out 24 times per second. Hence, we have an integration-ramp consisting
of 48 points. Although decreasing the number of data points on which
the fit is based might sound like a loss in signal-to-noise, one
should bear in mind that those points not used now, made a large
contribution to the variance with respect to the fitted slope because
of their non-linear behaviour. As a result, the error bars actually
decrease significantly.}. We first tested this procedure on the
ISO-SWS primary calibrator $\alpha$~Bootis and found that it does not
introduce any other artefacts.
%
%
\begin{figure}
\centering
  \resizebox{\hsize}{!}{\includegraphics{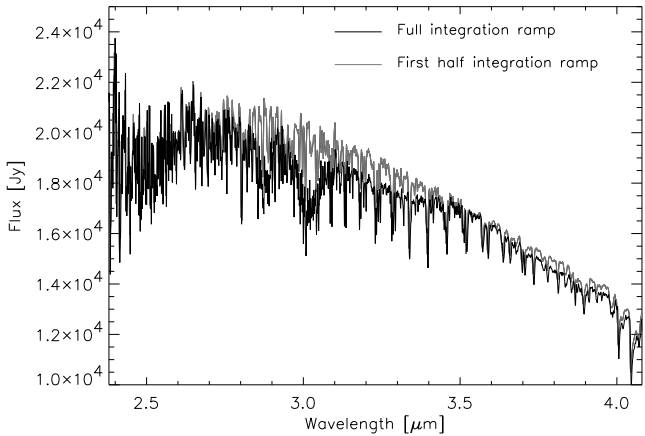}}
  \caption{
  The final band 1
  spectra of \aori: either based on uncorrected data (black) or on
  data corrected for presaturation (grey). The spurious spectral
  feature around $3.0 \mu$m has disappeared.
  }
  \label{fig:aori_corrected}
\end{figure}  

A comparison between the 2 spectra, the original one and a corrected
one, after reduction, is shown in Fig.~\ref{fig:aori_corrected}: the
spurious spectral feature around $3.0 \mu$m has disappeared.

\paragraph{Memory effects}

\begin{figure}
\centering
  \resizebox{\hsize}{!}{\includegraphics{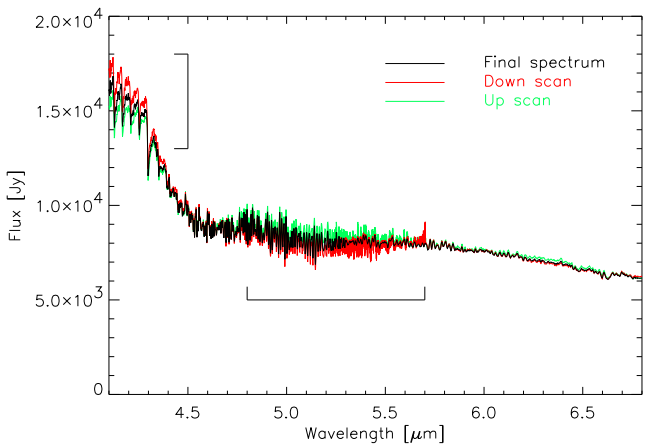}}
  \caption{Up scan (green), down scan (red) and final spectrum (black)
  in band 2a and 2b. Memory effects are strong at the blue side and at
  5.4~$\mu$m. The final flux may not be estimated correctly from 4.08
  to 4.3~$\mu$m and between 4.8 and 5.5~$\mu$m. We keep
  this in mind when comparing to the models.
  }
  \label{fig:memeffect}
\end{figure}

Band 2 (Si:Ga detectors, $4.08-12$~$\mu$m) is notoriously sensitive to
memory effects, especially for sources with a flux above 1000~Jy such
as \aori \citep{Kester2001}. The detector 'remembers' a previous
illumination and therefore predicts too large a flux for the current
observation. Since each sub band is scanned both from short to long
wavelengths and vice versa, it is possible to estimate the strength of
the memory effects. Fig.~\ref{fig:memeffect} shows up scan, down scan
and final version of the \aori spectrum. Strong memory effects are
indeed present. We must therefore keep in mind that the correction may
not be flawless from 4.08 to 4.3~$\mu$m and between 4.8 and
5.5~$\mu$m.

%
%

\paragraph{Absolute calibration}

Another peculiarity of the reduction of this observation concerns the
multiplicative\footnote{for bright sources like $\alpha$~Orionis, the
additive terms, such as the dark current, are negligible in comparison
to the multiplicative effects due to e.g. pointing errors, problems
with the spectral response function, etc.} factors we use to paste the
different sub bands together. The general strategy is the following:
band 1d ($3.02-3.52~ \mu$m) is believed to have the best absolute flux
calibration. Using the overlap between the different sub bands, it is
then possible to shift the other bands to the correct level, starting
from those adjacent to 1d and moving band-by-band to the red and
blue. Usually the factors used are quite close to unity, with
differences up to a few percent. However, for the spectrum of
$\alpha$~Orionis, several factors differ significantly from unity. We
remark that scaling factors closer to unity would be possible if we
assume the absolute calibration of band~1d to be too low
(see Sect.~\ref{sect:disc_aori}). Also the memory effects at 4.1~$\mu$m
could influence the scaling factor of band~2a by 5\%.
 For
completeness, we list these scaling factors in Table~\ref{table:factors}.
\begin{table}[th]
\centering
\caption{
 Multiplicative factors used to scale the different sub bands of the
 \aori spectrum to the absolute flux level when the latter is either
 determined by the flux level in band 1d (3rd column) or by requiring
 the average of the scaling factors to be equal to 1 (4th column). The
 large deviations from unity in the 1d-referenced scaling suggest that
 the absolute flux in band~1d might be underestimated. Further
 evidence supporting this hypothesis is given in
 Sect.~\ref{sect:disc_aori}. The large jumps between some bands can be
 (partially) attributed to the use of three different filters and
 apertures and 4 different detector types. 
 }
\vspace{2ex}
\begin{tabular}{l|ccc}
\hline
\hline
 Band    & $\lambda$ [$\mu$m] & 1d & mean$=$1  \\
\hline 
1a       & $2.38-2.60$  & 0.97 & 1.15 \\
1b       & $2.60-3.02$  & 0.98 & 1.16 \\
1d       & $3.02-3.52$  & 1.00 & 1.19 \\
1e       & $3.52-4.08$  & 0.97 & 1.15 \\
2a       & $4.08-5.30$  & 0.78 & 0.93 \\
2b       & $5.30-7.00$  & 0.80 & 0.95 \\
2c       & $7.00-12.0$  & 0.66 & 0.79 \\
3a       & $12.0-16.5$  & 0.71 & 0.84 \\
3c       & $16.5-19.5$  & 0.75 & 0.89 \\
3d       & $19.5-27.5$  & 0.76 & 0.90 \\  
\hline
\end{tabular}
\label{table:factors}
\end{table}

\subsection{Interferometry}

\paragraph{Literature data} 
\aori has been studied extensively with interferometric techniques in
the Near- and Mid-IR. In this work, we use 3 sets of IR observations
that were already published: (1) broad band FLUOR observations at
2.2~$\mu$m first presented in \cite{RSG2004}, (2) TISIS L band
observations at 3.8~$\mu$m \citep{Chagnon2002} and (3) ISI narrow-band
Mid-IR observations at 11.15~$\mu$m \citep{Weiner2003}.

The K band data cover molecular bands of \water and CO, the L band
data those of \water and SiO. These molecules
may very well be present in the {\sl molsphere} around
$\alpha$~Orionis. We note that,
if the extra molecular layers are optically thin, both bandpasses
remain photosphere-dominated and are therefore well suited for our
analysis. 

The mid-IR visibilities not only sample the photosphere+molecular
layers, but also the dust shell which emits strongly at 11~$\mu$m.  In
the case of $\alpha$~Orionis, the silicate emission of the dust shell
originates in a region far from the central object: the inner radius,
R$_{{\rm in}}$, is believed to be 0.5~arcsecond \citep{Sloan1993} or
even 1~arcsecond on the sky \citep{Danchi1994}, to be compared to
22~mas for the photospheric radius \citep{RSG2004}. The silicate
emission is therefore totally resolved at the spatial frequencies of
interest for the study of the central object. If one knows the flux
ratio $f=\frac{F_{star}}{F_{dust}}$, it is possible to renormalize the
observed visibilities allowing an independent study of the central
object without any other information on the dust shell. For
$\alpha$~Orionis, this flux ratio at 11.15~$\mu$m is well-determined
through other ISI observations \citep{Danchi1994,Bester1996,Sudol1999}
at very low spatial frequencies which sample the visibility curve at
the point where the dust shell is being resolved. The derived values
for the flux ratio $f$ range from 55 to 65~percent.  

\paragraph{New data}
\aori was observed in the N-band with the Mid-IR interferometric
instrument MIDI \citep[e.g. ][]{Leinert2003a} on the VLT-I
\citep[e.g. ][]{Glindemann2000} in Science Demonstration Time (SDT) on
November 8 and 10, 2003. The interferometric data will be presented
and analysed in a forthcoming paper, together with additional
MIDI observations scheduled for autumn 2005 which will cover a wider
range in spatial frequencies. 

\paragraph{MIDI single-dish spectra}

As shown in \cite{Chesneau2005}, it is
also possible to obtain good quality N-band spectra from MIDI
observations: for calibration purposes, single-dish spectra are
observed with both UT telescopes separately just prior or after the
interferometric observations. We searched these spectrally dispersed
1-D images (0.54~arcsec wide slit) for extended emission from the
detached dust shell but found none. Spectra were constructed by
integrating the images in the spatial direction with or without mask,
which did not yield significant changes in spectral shape. We use here
the spectra constructed with mask because of their higher S/N. They
were calibrated using the MIDI spectra of calibrator stars (HD~39400,
K1.5IIb and 17~Mon, K4III) observed in between the Betelgeuse
observations. The final averaged spectrum is shown in
Fig.~\ref{fig:midi_spectra}.


\section{The modelling}
\label{sec:modelling}

The aim of this investigation is to simultaneously model IR
spectroscopic and interferometric data in order to get a better
understanding of the physical processes at work between photosphere
and dust shell. Since we expect both the extended atmosphere and the
dust shell to be optically thin, it is crucial that we treat every
source of radiation properly. We can for example not neglect the
strong molecular bands already present in the photospheric spectrum of
a late-type supergiant: clearly not all molecular features seen in the
IR originate in ``extra'' molecular layers.

Hydrodynamical models of oxygen-rich late-type stars are not yet in a
stage where they can provide an accurate reproduction of observations,
and the general consensus is that we are still missing some fundamental
knowledge on the major processes, such as the dust condensation
sequence \citep[e.g. ][]{Tielens1990}. Moreover, these models are
currently built only for low-mass (Mira and C) stars, and their
applicability to supergiants is not clear. Therefore, we choose not to
use such a self-consistent model, but instead construct a
semi-empirical model as described below.

We will divide our model into 3 parts: (1) the hydrostatic
photosphere, (2) a region of extra molecular layers and (3) the dust
shell. The approximation of the extended atmosphere, i.e. the part
containing the extra molecular layers, with a discrete instead of a
continuous density, temperature and composition distribution is
supported by the first generation of hydrodynamical models for AGB
stars which predicts these distributions to be strongly radially
peaked \citep[e.g. ][]{Woitke1999,Hoefner2003}.

For each part, we have used the most up-to-date models
available. They are described below.

\subsection{The photosphere}
\label{sec:photosphere}

To model the photosphere of this supergiant, we use the {\sc
sosmarcs} code, version May 1998, as developed by
\citet{Gustafsson1975,Plez1992,Plez1993} at the university of Upssala,
Sweden. This code is specifically developed for the modelling of cool
evolved stars, i.e. a lot of effort was put into the molecular
opacities, and they allow for the computation of radiative transfer in
a spherical geometry. The line lists ---relevant for the IR part of the
spectrum--- used are those of \citet{Goldman1998} for OH,
\citet{Goorvitch1994} for CO, \citet{Langhoff1993} for SiO and those
of \cite{PartridgeSchwenke1997} for H$_{2}$O. This code solves
the radiative transfer equation in a hydrostatic LTE environment with
an ALI\footnote{ALI $=$ Approximate Lambda Iterator} method \citep[e.g.][]{Nordlund1984}. Opacities are treated
with the opacity sampling (OS) technique at 153910 wavelength points, which
guarantees a good sampling of both molecular and continuum opacity
sources. This OS grid offers a good compromise between computational speed
and accuracy of the atmospheric structure of the model. In
the interpretation of the similarities and discrepancies between
observed and synthetic spectrum, we make use of the study of the
influence of the stellar parameters on a synthetic spectrum performed
by \citet{Decin2000}. The initial values of the model parameters (which
are T$_{\rm{eff}}$, $\log{g}$, mass, the abundances of C, N, and O, the
microturbulent velocity, $\xi_{\rm{t}}$ and the metallicity [Fe/H])
were determined in great detail for $\alpha$~Orionis by
\cite{Lambert1984}.  They are listed in Table
\ref{paralambert}.
%
%
\begin{table}[ht]
 \begin{center}
 \caption{
 The stellar/atmospheric parameters of $\alpha$~Orionis, according to
 \citet{Lambert1984}, for a temperature of 3800\,K (their own estimate)
 and for a temperature of 3600\,K (as suggested by other temperature
 determinations).
 }
 \vspace{3ex}
 \begin{tabular}{lrr}
\hline
\hline
 parameter             & value            & value          \\
 \hline
 $\rm{T}_{\rm{eff}}$   & 3800\,K          & 3600\,K        \\
 $\log{g}$             & 0.0              & 0.0            \\
 Mass                  & 15 M$_{\odot}$   & 15 M$_{\odot}$ \\
 $\epsilon (\rm{C})$   & $8.41$           & $8.29$         \\
 $\epsilon (\rm{N})$   & $8.62$           & $8.37$         \\ 
 $\epsilon (\rm{O})$   & $8.77$           & $8.52$         \\
 $\xi_{\rm{t}}$        & 4 km/s           & 4 km/s         \\
 $[\rm{Fe}/\rm{H}]$    & 0.00             & 0.00           \\
 \hline
 \end{tabular}
 \label{paralambert}
 \end{center}
\end{table}

For $\alpha$~Orionis, masses derived from theoretical
Hertzsprung-Russell diagrams range from 15\,M$_{\odot}$
\citep{LambIbenHoward1976, CloutmanWhitaker1980} to 30\,M$_{\odot}$
\citep{StothersChin1979}. Adopting a distance of 131 parsec
\citep[Hipparcos,][]{Perryman1997} and a photospheric angular
diameter of 43\,mas yields a stellar radius of about
645\,R$_{\odot}$. This result confirms the value for the gravity found
by \cite{Lambert1984}, derived from the fact that neutral and ionized
lines should yield the same abundances. We stress the quality of this
determination of the surface gravity because photospheric molecular
bands, and especially those of water, are highly sensitive to the
surface gravity \citep{Decin2000}.

\begin{figure}[t]
 \vspace{2ex}
 \begin{center}
 \resizebox{\hsize}{!}{\includegraphics{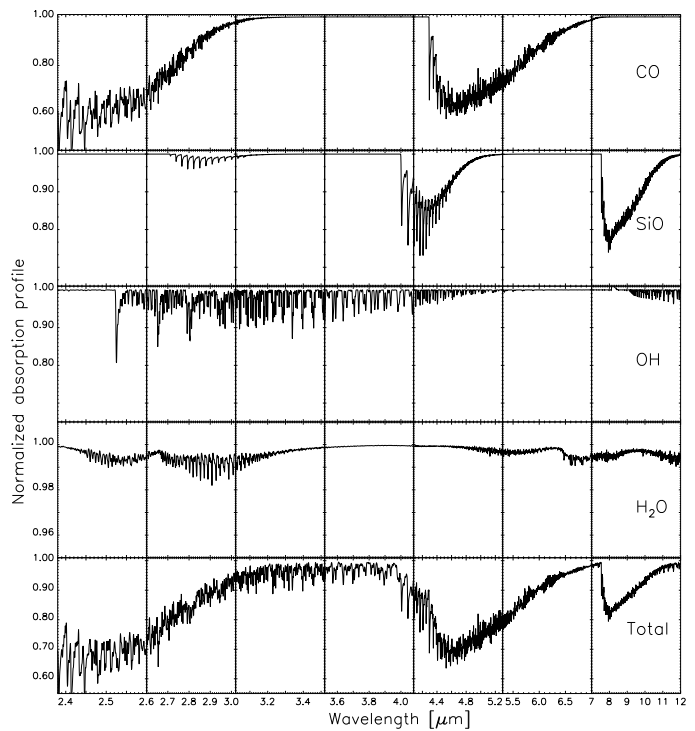}}
 \vspace{6ex}
 \caption{
 Main contributions to the total photospheric absorption in the
 atmosphere model with the stellar parameters from \cite{Lambert1984}
 at 3600~K. CO and SiO are clearly dominant, while there are only
 minor traces of absorption by water. The relevant line lists used are
 those of \citet{Goldman1998} for OH, \citet{Goorvitch1994} for CO,
 \citet{Langhoff1993} for SiO and those of Ames
 \citep{PartridgeSchwenke1997} for H$_{2}$O. 
\label{fig:molecules}
 }
\end{center}
\end{figure}

The molecules responsible for absorption in the ISO-SWS wavelength
region are primarily CO, H$_{2}$O, OH and SiO. The contribution of
these molecules to the total absorption, when adopting the stellar
parameters of \citet{Lambert1984}, is shown in
Fig.~\ref{fig:molecules}.

%
%

\subsection{The molecular layers}

Several approaches are possible when modelling {\sl molspheres}. They
range from plane-parallel slabs placed in front of the star
\citep[e.g.][]{Yamamura1999,Matsuura2002,Cami2002,VanMalderen2003}
over infinitesimally thin spherical shells \citep{Mira2004} to
extended spherical layers with temperature and density distributions
\citep{Mennesson2002,Ohnaka2004b,Ohnaka2004}. 

Slab models are not suitable for our purposes: the molecular layers
are expected to be fairly optically thin, in which case the
slab-approximation is far too rough (in a first order because these
models do not take into account that the visible column density is
twice as high next to the stellar disk as it is in front of it, and to
a second order because they miss sphericity effects which are
important for large scale heights). Both the resulting spectra and the
resulting visibilities are therefore not realistic for optically
thin\footnote{This does not take away their merit in the modelling of
the thick molecular layers surrounding Mira-like stars, where the
column density and consequently also the opacity are so large that the
backside of the shell is not visible and the spatial intensity
distribution is rather UD-like.} {\sl molspheres}.

On the other hand, the shocked nature of these layers most probably
implies that they are geometrically quite thin and that therefore
temperature and density gradients within the shell can be
neglected. Moreover, the observational data are at the moment too
limited (in both spectral and spatial resolution) to constrain these
supplementary parameters. 

\begin{figure}
\centering
  \resizebox{9cm}{!}{\includegraphics{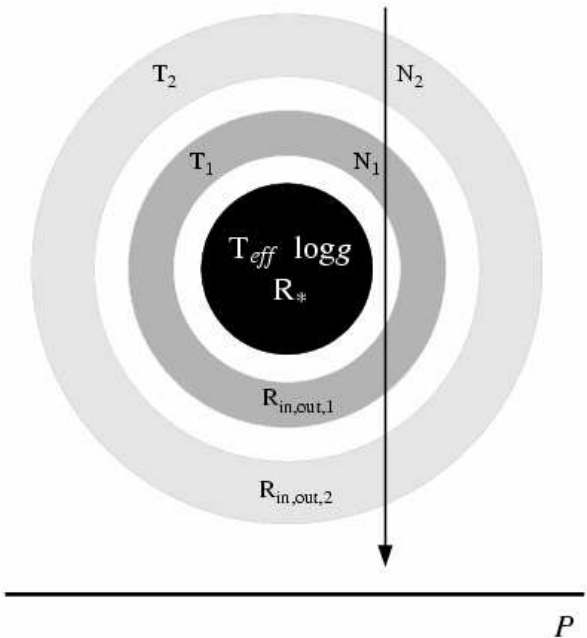}}
  \caption{The structure of the models presented here: a photospheric
  model which includes limb darkening and the relevant molecular
  opacity, surrounded by layers of possibly mixed composition, each
  with its temperature, density, inner radius and outer radius. For
  the silicate feature modelling, the resulting spectrum is fed into
  the dust radiative transfer code {\sc modust} \citep{Bouwman2000}.
  }
  \label{fig:model_structure}
\end{figure}

Consequently, we opt for isothermal, spherical layers with a finite
extent but no density distribution, which are characterized by their
temperature, composition, column density for each molecule, inner
radius and outer radius. Fig.~\ref{fig:model_structure} visualizes the
structure of the models used.

%
%

\subsubsection{Opacities}

Opacities were calculated for H$_2$O, SiO, CO, and OH at
temperatures ranging from 500 to 2500~K, with 100~K increments.

For H$_2$O, we used the {\sc nasa ames} line list and partition
function \citep{PartridgeSchwenke1997}, which is at present the most
complete list, including more than 300 million lines. However, for
computational reasons, we included only lines with $\log{gf -
\chi\theta} \ge -9$, with $g$ the ground level statistical weight, $f$
the oscillator strength, $\theta = 5040/3500$ and $\chi$ the
excitation energy in eV. The consequences of neglecting the weaker
lines on the final opacity are below 1 percent for a spectral
resolution of 300 \citep{VanMalderen2004}. Differences between the
available line lists for \water (e.g. {\sc ames, hitemp, scan}) are
significant. The {\sc hitemp} line list is aimed at temperatures of
about 1000~K, and might not be valid for the very high temperatures of
the water around Betelgeuse.  \cite{Decin2003d} found the best
reproduction of observed water spectra in the ISO-SWS spectra of M
giants with the {\sc ames} line list, for which reason we used it here
as well.

CO, OH and SiO present far less difficulties than water. We used the
line list of \cite{Goorvitch1994}, \cite{Goldman1998} and
\cite{Langhoff1993} respectively. These are the most complete lists to
date for astrophysical applications \citep{Decin2000}. The polynomial
expansion of the appropriate partition function was taken from
\cite{Sauval_Tatum_1984}.

As microturbulent velocity we use 3~km~s$^{-1}$ as suggested for cool
giants by \cite{Aringer_2002}.

\subsubsection{Radiative transfer}

We compute emerging intensities for a grid of linearly spaced impact
parameters $p$ (256 in total), i.e, we construct a 1-D intensity
profile $I(\lambda,p)$, going from the center of the disk to the
edge. From this intensity profile, both the resulting spectrum and the
visibilities can be calculated. For more details on the calculation of
the radiative transfer, we refer to \cite{Thesis}.

\begin{figure}[ht]
\centering
 \resizebox{\hsize}{!}{\includegraphics{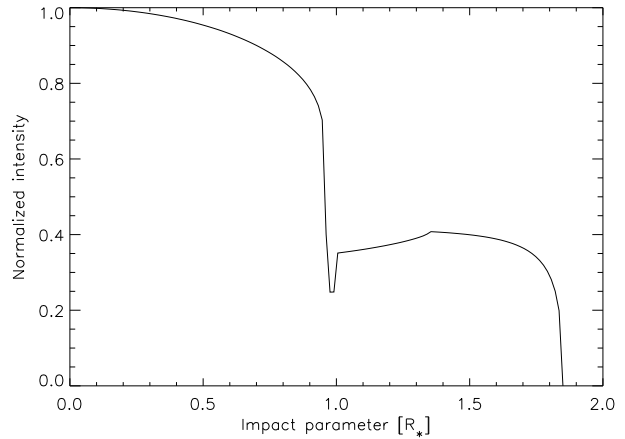}}
 \caption{
An example of an intensity profile at 2.9~$\mu$m for our {\sc marcs}
photosphere, surrounded by a single layer with R$_{\rm{in}}=1.35
\rm{R}_{\star}$, R$_{\rm{out}}=1.85 \rm{R}_{\star}$, a temperature of
2300~K and a H$_2$O column density of $5 \times
10^{20}$~cm$^{-2}$. The dip in the IP at $\sim 1$\,R$_{\star}$ is an
artefact due to the assumption that the central star is optically
thick also in the limb.
 }
\label{fig:ip_example}
\end{figure}

An example of such an intensity profile for a single layer around a
limb-darkened central star is shown in Fig.~\ref{fig:ip_example}.
%
%
From the theoretical wavelength-dependent intensity profile, we
calculate the spectrum by integration over the emitting surface and
the monochromatic visibility by a Hankel transform
\citep{Hanbury-Brown74} of the intensity profile.  This monochromatic
visibility is then convolved with the bandpass of the observations for
the comparison model vs. observations. For wide-band data close to and
beyond the first null, bandwidth effects become important: because of
the wavelength dependence of the spatial frequency at which we
observe, one wide band visibility measurement actually covers a range
in spatial frequencies, but is assigned to the spatial frequency
corresponding to the effective wavelength of the filter. We follow the
approach of \cite{RSG2004} and take this effect into account by
averaging properly weighted squared visibilities.

\subsection{The dust shell}

The dust shell is modelled using the proprietary spherical radiative
transfer code {\sc modust}\\ \citep{Bouwman2000,Bouwman2001}. Under
the constraint of radiative equilibrium, this code solves the
monochromatic radiative transfer equation from UV/optical to
millimetre wavelengths using a Feautrier type solution method
\citep{Feautrier1964, Mihalas1978}. The code allows to have several
different dust components of various grain sizes and shapes.  


\section{Comparison with observations}
\label{sec:comparison_aori}

When confronting the synthetic spectrum based on our {\sc marcs} model
atmosphere with the ISO-SWS spectrum of \aori
(Fig.~\ref{fig:vglallbands}), it is clear that our model makes a
good reproduction of the global spectral shape. This shape is quite
different from that of a blackbody: clearly, the major sources of
opacity have a strong influence on the spectral
appearance in the IR, which is well reproduced by the {\sc marcs}
model atmosphere. However, several differences can be observed: 
\begin{itemize}
\item between 2.38 and 3.5\,$\mu$m the model strongly overestimates the flux,
\item from 4.0 to 4.2\,$\mu$m the flux is underestimated,
\item around 5~$\mu$m, there are extra absorption features in the ISO-SWS spectrum, 
\item also from 6\,$\mu$m onward there is extra absorption, and
\item from 8.5~$\mu$m on-wards, the observed spectrum is dominated by
silicate emission.
\end{itemize}

The latter property clearly divides the spectrum in a part without
dust contribution, and a part which will require a modelling of the
dusty outflow. We are therefore motivated to separate at first the
analysis of the near-IR part of the spectrum from that of the mid- to
far-IR wavelength regime. 

\subsection{The near IR}
We start with a study of the near-IR wavelength regime, where
photospheric emission dominates, but several discrepancies require
further investigation, as noted above.  
\subsubsection{Comparing {\sc marcs} with ISO-SWS} 
Unfortunately, the first three
discrepancies coincide with wavelength regions not free of calibration
problems: the first one corresponds to that affected by
presaturation. We are however confident that this issue was dealt with
in an appropriate way
as discussed in Sect.~\ref{sec:obs:iso}, and that the discrepancy seen
is of a physical nature. As to the 2nd region: this is the blue side
of band~2a, which shows strong memory effects as shown in
Sect.~\ref{sec:obs:iso} and Fig.~\ref{fig:memeffect}. The same is true
for the 3rd region, but the discrepancy is much wider than the
wavelength band affected by memory effects
(cf. Fig.~\ref{fig:memeffect}). Since a correction for memory effects
was applied, we can at this point not rule out the possibility that
these are also true discrepancies between model and reality.

%
%
\begin{figure}[t]
\begin{center}
 \resizebox{\hsize}{!}{\includegraphics{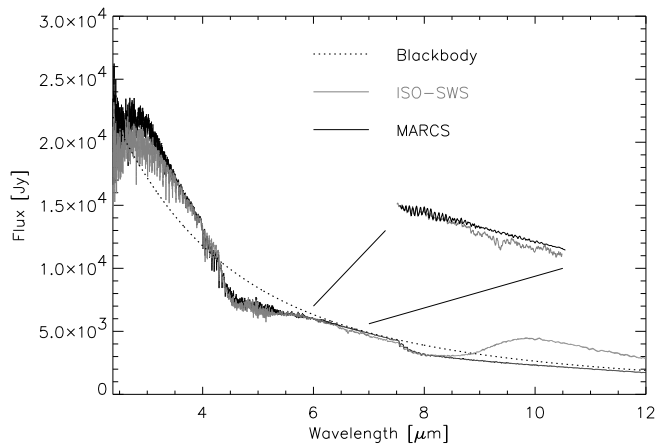}}
 \caption{
     A comparison between the ISO-SWS-spectrum (grey) and our
     synthetic {\sc marcs} spectrum (black). A blackbody at T=3600\,K is
     shown by the dotted line for comparison. Clearly, our model
     can reproduce the global spectral shape very well. Nevertheless,
     there are some interesting discrepancies.
 }
 \label{fig:vglallbands}
\end{center}
\end{figure}

From Fig.~\ref{fig:molecules}~and~\ref{fig:vglallbands}, it is clear
that the first and fourth discrepancy, i.e. the strongest ones, can be
due to unpredicted absorption by water, while the second one appears
to correspond to SiO. Some contribution of extra CO absorption might
also be present in the 2.38 and 5~$\mu$m discrepancies.

However, for the parameters being studied, the model predicts almost
no absorption by water: the atmospheric temperature is too high,
except for the outermost layers. Lowering the effective temperature is
not an option, because it is well constrained by the apparent diameter
and the bolometric flux \citep[e.g.][]{RSG2004}.  Using a higher
oxygen abundance does not work either since a very large increase would
be required, which then causes far too strong spectral features by OH
and SiO. Also an attempt at increasing the \water absorption with a
higher surface gravity fails: a physically impossible value of
$\log{g} = 2$ would be required (see Sect.~\ref{sec:photosphere} for a
discussion on the determination of the gravity).  

Concerning the excess emission by SiO at 4.2~$\mu$m: since the
SiO band at 8~$\mu$m is well predicted by the model, it is most likely
only a memory effect and therefore not real.

We conclude that we see clear extra absorption by \water which cannot
be predicted by the hydrostatic {\sc marcs}  model with reasonable input parameters. The
problem thus appears not to be in the input parameters but in the basic
assumptions. Three of these are not very likely to hold in the
atmosphere of a variable supergiant: LTE, homogeneity and hydrostatic
equilibrium. 

LTE is most definitely violated in the outermost layers of the
photosphere where the density is low \citep[see
  e.g. ][]{Ryde2002}. However, non-LTE effects are generally only seen
in high-resolution spectra \citep[e.g. ][]{Ayres1989}, and are not
believed to create pseudo-continuous effects such as seen here in
$\alpha$~Orionis. 

The second assumption, homogeneity, is probably also violated in the
atmosphere of Betelgeuse. \cite{Freytag2002} calculated convection
models for Betelgeuse and found large convective cells with strong
temperature gradients between the border regions and the
center. Possibly, more molecules, including H$_2$O, could form there
too.

The last assumption has been well studied in the context of Mira
stars. Hydrodynamic models are essential in the modelling of observed
IR spectra of these pulsators \citep{Woitke1999,Hoefner2003}.
Although the amplitude of the visual variations of \aori is small
compared to that of a typical Mira, i.e. about 0.25~mag in V
\citep{Gray2000} vs. 3~mag for e.g. T Cep \citep{VanMalderen2003}, the
low surface gravity and high luminosity may help the levitation of the
upper layers. As for Mira stars, the temperatures in this levitated
matter can be low enough for molecules to form. If shocks are formed,
by outward moving material running into infalling material, high
molecular density shells will result.   

Extra molecular layers are thus an attractive solution for the
observed discrepancies. We will investigate this hypothesis below, but
let us first remark that the angular diameter on the sky required to
fit the ISO-SWS spectrum is only 43.6~mas (limb-darkened diameter of
the $\tau_{\rm{ross}=10^{-7}}$~layer), to be compared to an observed
LD diameter of 45.6~mas (obtained by fitting L band visibilities
computed from our {\sc marcs} model to the TISIS
observations\footnote{The agreement between photosphere model and
ISO-SWS spectrum is excellent across the L band. There should be no
contribution of extra layers at these wavelengths}). This confirms the
hypothesis formulated in Sect.~\ref{sec:obs:iso} that the absolute
calibration of the ISO-SWS spectrum based on band~1d has caused an
underestimation of the actual flux, resulting in both the large
deviations from unity for the scaling factors presented in
Table~\ref{table:factors} and the discrepancy in photometric and
interferometric diameter reported here. The interferometric diameter
would suggest the actual flux to be about 10\% higher than predicted
by band 1d (cfr. Table~\ref{table:factors}).

\subsubsection{Adding an \water layer} 
Fitting the ISO-SWS spectrum with a photosphere+water-layer
model, by a $\chi^2$ minimization in the grid presented in
Table~\ref{tab:aori_layer_grid}, we find the best fit for a layer of
1750~K, with an outer radius of 1.45~R$_{\star}$ and an H$_2$O column
density of $2 \times 10^{19}$~cm$^{-2}$.

\begin{figure}[t]
\centering
 \resizebox{\hsize}{!}{\includegraphics{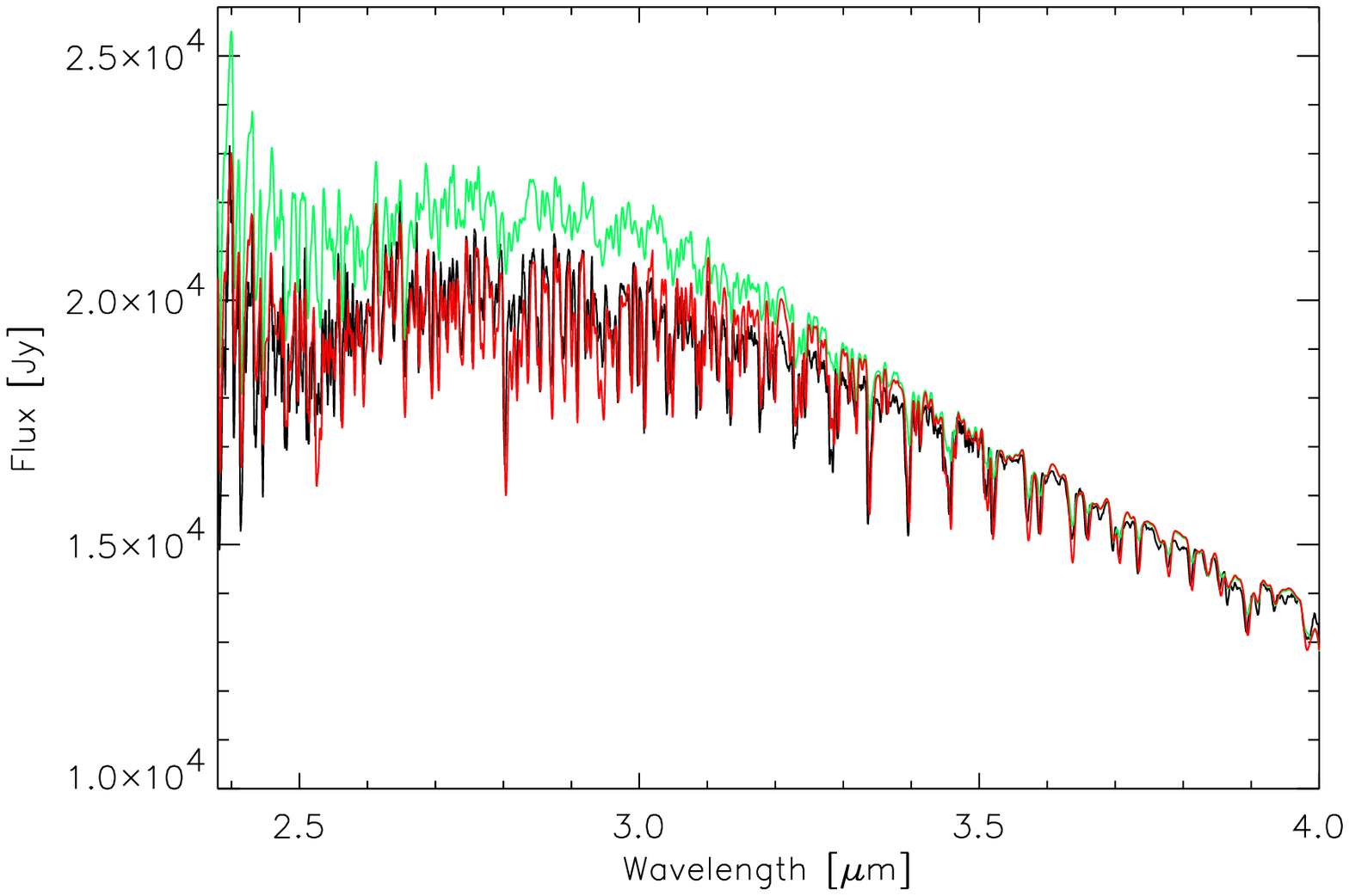}}
 \resizebox{\hsize}{!}{\includegraphics{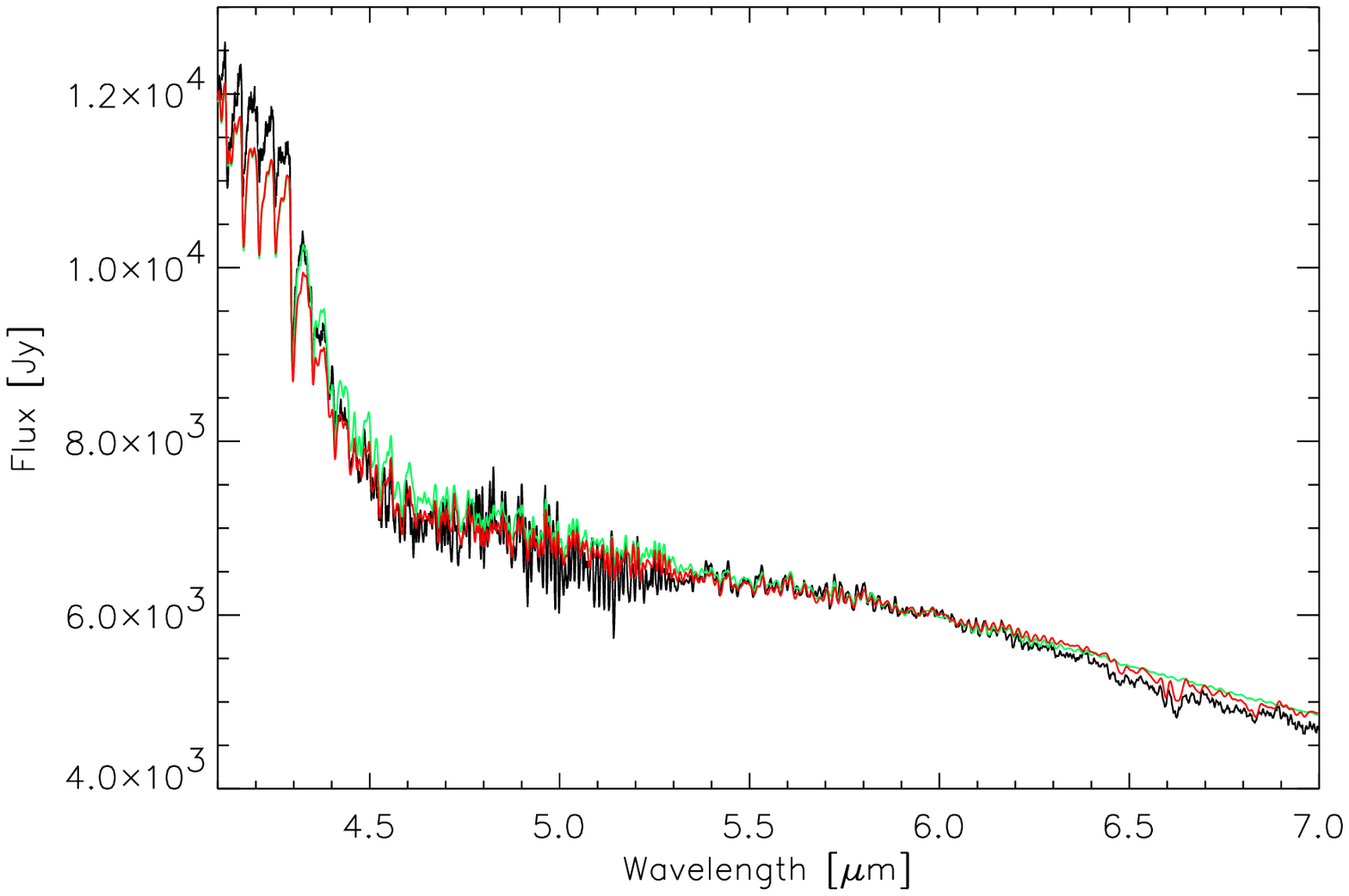}}
 \caption{
 The ISO-SWS spectrum (black) is compared to the {\sc marcs}
 photosphere model (green), and the {\sc marcs} photosphere model
 embedded in a molecular layer (red) at 1.45~R$_{\star}$, with a
 temperature of 1750~K and H$_2$O, CO and OH column densities of $2
 \times 10^{19}$~cm$^{-2}$, $2 \times 10^{21}$~cm$^{-2}$, and $5
 \times 10^{20}$~cm$^{-2}$ respectively. The improvement is
 significant. The problems at 4.2 and 5~$\mu$m can be attributed to
 memory effects in the ISO-SWS observation (see
 Fig.~\ref{fig:memeffect}).
 }
\label{fig:aor_iso_1_2ab}
\end{figure}

A few discrepancies remain, and they can be identified as due to CO
(at 2.4 and 4.5 $\mu$m) and OH (several strong lines between 3 and 3.5
$\mu$m). The fit to the ISO-SWS spectrum can be improved some more
by adding $2 \times 10^{21}$~cm$^{-2}$ of CO and $5 \times
10^{20}$~cm$^{-2}$ of OH.  The resulting synthetic spectrum is shown
in Fig.~\ref{fig:aor_iso_1_2ab}. 
%
%
 \begin{table}[t]
 \begin{center}
 \caption{
 The grid of layer parameters in which we search for a best fit to
 both the ISO-SWS spectrum and the interferometric data. 
 }
 \vspace{3ex}
 \begin{tabular}{lccc}
\hline
\hline
 parameter                    & min. val. & max. val. & stepsize  \\
 \hline
 R$_{\rm{out}}$ [R$_{\star}$] & 1.20      & 1.50      & 0.05      \\
 Temp. [K]                    & 1500      & 2500      & 100       \\
 Col. dens. H$_2$O [cm$^{-2}$]& $6 \times 10^{18}$      & $4 \times 10^{20}$      & $\times$ 2 \\
 Col. dens. CO [cm$^{-2}$]    & $1 \times 10^{20}$      & $1 \times 10^{24}$      & $\times$ 10\\
 Col. dens. OH [cm$^{-2}$]    & $1  \times 10^{20}$      & $1 \times 10^{24}$      & $\times$ 10\\
 Col. dens. SiO [cm$^{-2}$]   & $1  \times 10^{20}$      & $1 \times 10^{24}$      & $\times$ 10\\
\hline
 \end{tabular}
 \label{tab:aori_layer_grid}
 \end{center}
\end{table} 

\subsubsection{Comparison with the near-IR interferometric data} 
\begin{figure}[t]
\centering
 \resizebox{\hsize}{!}{\includegraphics{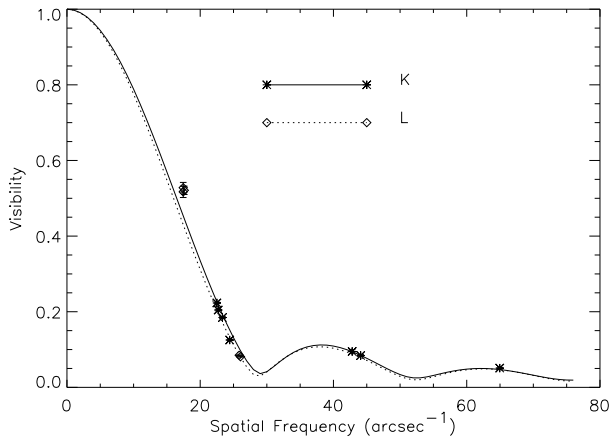}}
 \caption{
 Observed K and L band visibilities are compared to our best single
 layer model. The match is excellent for a photospheric LD diameter of
 45.6~mas. The very discrepant points at low spatial frequency are
 TISIS L band observations, which might be corrupted by a poor
 subtraction of the thermal background.
 }
\label{fig:aori_interfero}
\end{figure}

From this model, we can compute K and L wide band
visibilities, to be compared to the FLUOR (K) and TISIS (L) observations. The
result is shown in Fig.~\ref{fig:aori_interfero}. The agreement is
convincing for a photospheric limb-darkened diameter of 45.6~mas. In
fact, the layer opacity in the FLUOR and TISIS bandpasses is so low
that there is no noticeable difference in visibility curve between pure
photosphere and photosphere+layer model.  
%
%
The derived photospheric diameter is slightly larger than
determined from the same data by \cite{RSG2004}, but this is due
to the extent of our stellar photosphere model: the outermost layer of
our model corresponds to R$_{\tau_{\rm{ross}}=10^{-7}}$. This
translates into a ``tail'' on the intensity profile which is not
present in the analytical profile used by \cite{RSG2004} and which
amounts to about 4 percent of the total stellar diameter. We conclude
that the photospheric diameter of our best-fitting photosphere+layer
model is in good agreement with the one found by \cite{RSG2004},
i.e. that the molecular layer does not change the apparent size in the
near-IR at low spectral resolution.

\subsection{The mid IR}
\subsubsection{Excess emission and a larger apparent diameter}
\begin{figure}[t]
\centering
 \resizebox{\hsize}{!}{\includegraphics{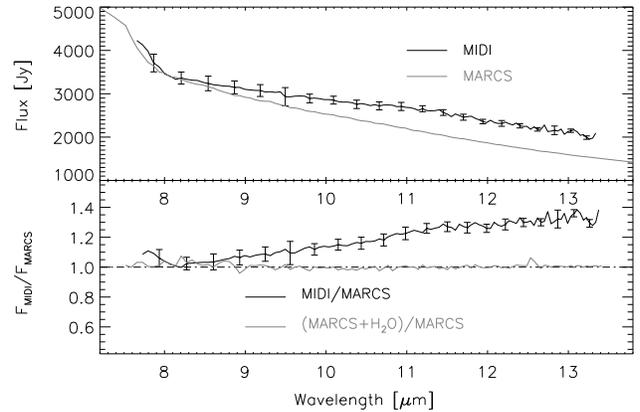}}
 \caption{
 Average MIDI N-band spectrum compared to the {\sc marcs} prediction for the
 photospheric emission. Upper panel: the calibrated MIDI spectrum and
 the {\sc marcs} photospheric spectrum. There is no trace of olivine emission in the MIDI slit, which is
 0.54\,arcsec wide, confirming that the olivine dust is indeed more
 than 10\,R$_{\star}$ out. Lower panel: the MIDI spectrum divided by
 the {\sc marcs} spectrum. There is excess
 emission within the PSF which increases with wavelength up to about
 $11-12$~$\mu$m and then levels out. 
 }
\label{fig:midi_spectra}
\end{figure}

\begin{figure}[t]
\centering
 \resizebox{\hsize}{!}{\includegraphics{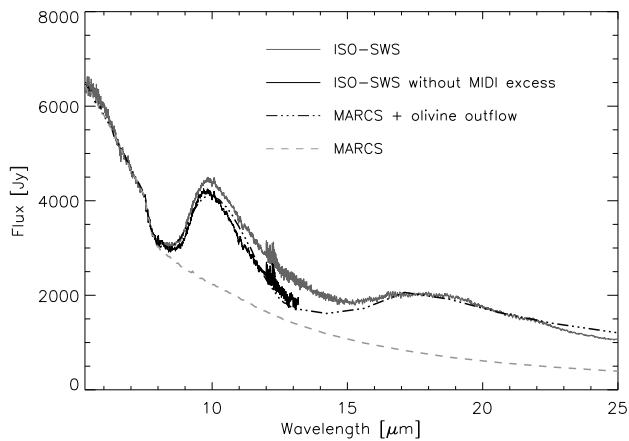}}
 \caption{
 The ISO-SWS spectrum of \aori (black) in the mid-IR is dominated by
 the amorphous olivine resonances at 9.7 and 18~$\mu$m
 respectively. If the dust features are modeled with an olivine
 outflow (parameters listed in Table~\ref{table:olivine_params}), a
 large discrepancy from $8-17$~$\mu$m suggests another source of
 emission/opacity. Indeed, when we subtract from the ISO-SWS spectrum the
 excess emission found in the MIDI N-band spectrum, the agreement is
 far better (solid line).  
 }
\label{fig:olivine_fit}
\end{figure}

The mid-IR part of the
ISO-SWS spectrum is dominated by the Si-O stretching and O-Si-O
bending resonances in amorphous olivines at 9.7 and 18~$\mu$m
respectively (Fig.~\ref{fig:olivine_fit}). The MIDI N-band spectra on
the other hand do not show such a 9.7~$\mu$m feature
(Fig.~\ref{fig:midi_spectra}). The slit used for the latter observations is
only 0.52~arcsec wide. This confirms the large inner radius
of the olivine dust shell as suggested by \cite{Sloan1993} who
found no silicate emission inside a region of about 0.5~arcsec, some
emission between 0.5 and 1~arcsec, and the better part of the silicate
emission even further out. This is also consistent with UKIRT mid-IR
images presented by \cite{Skinner1997} and with the inner radius
(1~arcsec) measured by \cite{Danchi1994} with the ISI interferometer.

%
%

Surprisingly, there is excess (non-silicate) emission within the PSF
of the MIDI N-band spectra which increases with wavelength up to about
11~$\mu$m and then levels out (Fig.~\ref{fig:midi_spectra}). That such
excess emission is present is also confirmed by a {\sc modust}
modelling of the dust shell based on the ISO-SWS spectrum: in
Fig.~\ref{fig:olivine_fit}, we see that the 2 features due to
amorphous olivine can be modelled quite accurately but that the flux
predicted in between is far too low. This problem cannot be solved
with other grain sizes, nor with another extent of the dust
shell. However, when subtracting the excess emission seen in the MIDI
spectrum from the ISO-SWS spectrum, we arrive at a far better
agreement between model and observations. Moreover, the excess seems
to decrease again beyond 13~$\mu$m, disappearing entirely at 17~$\mu$m.

\begin{table}[t]
 \begin{center}
 \caption{
  Dust shell parameters for the best fit to the ISO-SWS spectrum. The
  dust shell inner radius corresponds to the measured inner radius of
  \cite{Skinner1997} and \cite{Sloan1993}. The dust to gas ratio is
  taken from \cite{Knapp1980} and the outflow velocity from
  \cite{KnappMorris1985}. Grain shapes are a continuous distribution
  of ellipsoids (CDE).  
   \label{table:olivine_params}
  }
 \vspace{3ex}
 \begin{tabular}{lc}
  \hline
  \hline
  Parameter                    & Value \\
  \hline		    					  
\rule[0mm]{0mm}{5mm}$\dot{\rm{M}}_{\rm{dust}}$ & $6.3 \times 10^{-10}$~M$_{\odot}$ yr$^{-1}$\\
  v$_{\infty}$        & 15 km s$^{-1}$ \\
  Dust to gas ratio   & 0.0025 \\
  R$_{in}$            & 20 R$_{\star}$ \\
  R$_{out}$           & $\infty$ \\
  Composition         & Amorphous MgFeSiO$_4$ \\
  Grain size          &  $0.01-1$ $\mu$m \\
  Grain shapes        & CDE \\
 \hline
 \end{tabular}
 \end{center}
\end{table}

We remark that the dust mass loss rate we derive from a best fit to
the ISO-SWS spectrum ($6.3 \times 10^{-10}$~M$_{\odot}$ yr$^{-1}$) is
in good agreement with that found by \cite{Knapp1980},
\cite{KnappMorris1985}, \cite{Knapp1986} and
\cite{SkinnerWhitmore1987}. It is worth stressing that in the case of
this optically thin dust shell, it is important to take into account
the photospheric SiO band head at 8~$\mu$m when using the 9.7~$\mu$m
feature to determine the dust mass loss rate.

%
%

\cite{Weiner2003} present ISI observations (at 11.15~$\mu$m) performed
between November 1999 and December 2001 at spatial frequencies which
cover very well the first lobe of the visibility curve due to the
central object, i.e. the object inside e.g. the MIDI PSF. Together
with earlier observations at much smaller baselines which allow the
determination of flux ratio between central object and olivine dust
shell, these allow a diameter determination of the central object at
that wavelength. The flux ratio determined from these observations
ranges from $55-65$~percent \citep{Danchi1994,Bester1996,Sudol1999}
which is compatible with the flux from the central object in the MIDI
spectra. The diameter is about $55-60$~mas \citep{Weiner2003},
i.e. almost 1.5 times the size in the near IR. Moreover, no data in the
second lobe of the visibility curve are available, and limb-darkening
may therefore be significant, increasing the size of the central
object at 11.15~$\mu$m even more.   

\subsubsection{Explaining the mid-IR data}

In the previous section, we found that in the mid IR, \aori (without
the detached olivine dust shell) appears to be {\sl about 1.5 times as
large} as in the near IR. Moreover, we found {\sl excess emission}
from 8.5~$\mu$m onward, which increases smoothly with wavelength up
to about 11 or 12~$\mu$m, where the excess levels out at
$25-30$~percent. Beyond 17~$\mu$m there appears to be no longer an
excess.  There are 3 possible sources of extra opacity which
could be responsible for the observed excess and diameter increase:
(1) an extra molecular layer, (2) chromospheric emission and (3)
dust. Below, we investigate each of these possibilities.

\begin{figure}[t]
\centering
\resizebox{\hsize}{!}{\includegraphics{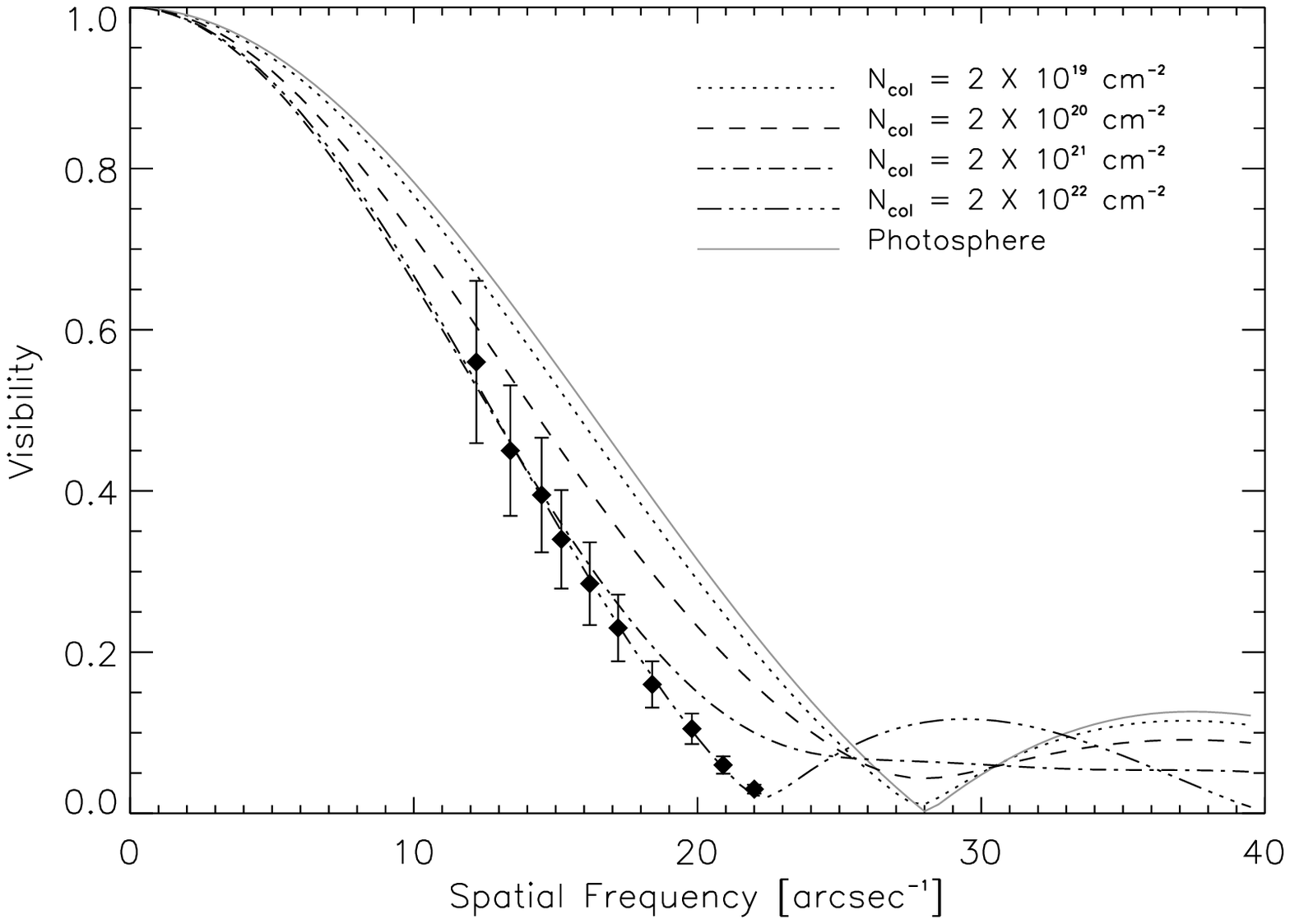}}
\resizebox{\hsize}{!}{\includegraphics{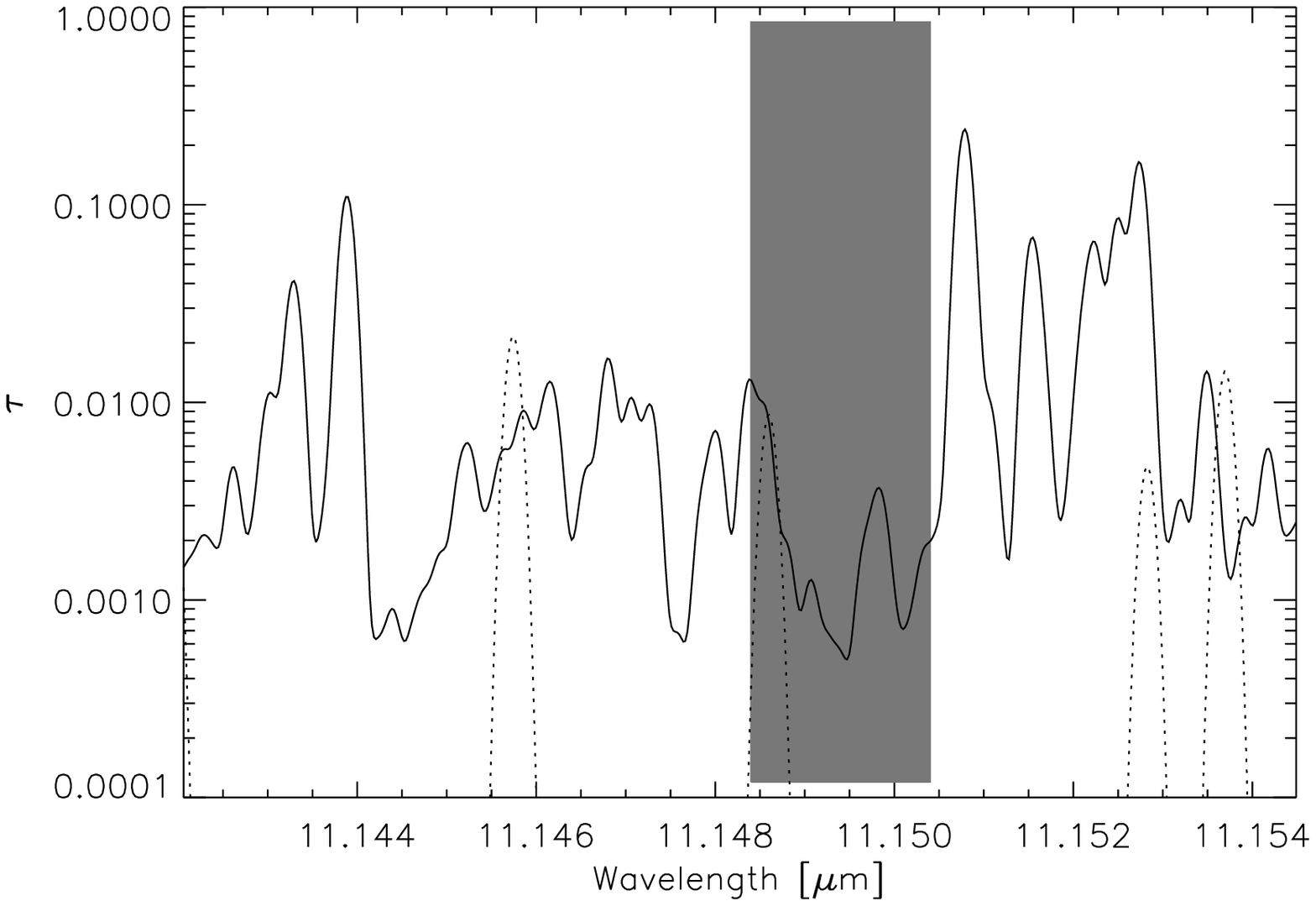}}
 \caption{
 Upper panel: Comparison of different models with the ISI observations
 (diamonds) at 11.15~$\mu$m made by \cite{Weiner2003}.  To match these
 data with an \water layer, we need a high H$_2$O column density of
 about $2 \times 10^{22}$~cm$^{-2}$, which is neither compatible with
 the near-IR spectrum, nor with the MIDI N-band spectrum. Lower
 panel: Opacity in the ISI bandpass (shaded region) of $2 \times
 10^{19}$~cm$^{-2}$ of \water (solid line) and $2 \times
 10^{21}$~cm$^{-2}$ of SiO (dotted line), both at a temperature of
 1750~K. The opacity profile is redshifted by 21~km\,s$^{-1}$, which
 is the radial velocity of \aori \citep{Wilson1953}.
 }
\label{fig:ISI_simulation}
\end{figure}

\paragraph{An extra molecular layer}
From an analysis of the near-IR part of the ISO-SWS spectrum, we found
evidence for an extra molecular layer at about 0.45~R$_{\star}$ above
the photosphere, containing mainly water. This putative layer is so
optically thin that it does not influence the apparent size in the K
and L bands. Since \water opacity is in fact even lower around
11~$\mu$m \citep[e.g.][]{VanMalderen2003}, we do not expect a
significant diameter increase at those wavelengths. Indeed, to
reproduce the ISI observations, the layer has to contain about $2
\times 10^{22}$~cm$^{-2}$ of water (upper panel of
Fig.~\ref{fig:ISI_simulation}), which is 1000~times more than what
we found from the near-IR part of the spectrum. Such a thick \water
layer is clearly not present: it would create optically thick water
bands in the near-IR as well, and those are not seen in the ISO-SWS
spectrum. Moreover, \water opacity has a minimum around 10~$\mu$m
which would yield a clear spectral signature in the MIDI N-band
spectrum (unless the layer is extremely thick and isothermal), which is
not observed. We conclude that \water opacity can not be responsible
for the observed diameter increase and excess emission.
%
%

However, there may be other molecular sources of opacity at
11.15~$\mu$m. As can be seen from Fig.~\ref{fig:molecules}, both SiO
and OH show spectral lines around 11~$\mu$m (TiO, which is not shown
in Fig.~\ref{fig:molecules}, does not have significant lines in the
mid-IR). Those of OH are well separated and not very numerous: our
line list \citep{Goldman1998} has no strong lines in the ISI
bandpass. The red wing of the opacity profile of hot SiO might cover
the ISI bandpass, but a very large column density would be required
to reach the required amount of opacity. This is not compatible with
the lack of an extra-photospheric SiO signature around 8~$\mu$m in the
ISO-SWS spectrum. The lower panel of Fig.~\ref{fig:ISI_simulation}
displays the opacity of \water and SiO in the ISI bandpass.  We
thus {\sl exclude a molecular cause} for the diameter increase and excess
emission is.

\paragraph{Chromospheric emission}

From UV and radio observations, \aori is known to have both a
hot chromospheric component and an ionized wind with fairly low
electron temperatures
\citep[e.g.][]{Basri1981,Gilliland1996,Skinner1997,Lim1998}.
\cite{Harper2001} present a model of the gas in the extended
atmosphere of \aori from $1-10$~R$_{\star}$, which matches the radio
data up to the mid IR. It consists of a weakly ionized wind with mainly
H$^-$ f-f opacity. The temperature distribution peaks at about 80~mas
(diameter) or 1.45 times the (mid-IR) ISI diameter. The maximum mean electron
temperature in their model is only about 4000~K, but since that is not
enough to explain the UV observations, they suggest an inhomogeneous
medium: the region around 1.45~R$_{\rm{ISI}}$ has an ambient temperature of
only 2000~K, but contains a multitude of chromospheric hot spots with
electron temperatures of the order of 10000~K. With an areal filling
factor (AF) of less than 0.25, this could reconcile the UV traces of a
hot chromospheric component with a mean electron temperature of only
4000~K. While the mean model predicts neither excess emission nor a
larger apparent size at 11.15~$\mu$m (Harper, private communication),
we investigate whether the suggested chromospheric hot spots
could be at the origin of the mid-IR excess and diameter increase.

\begin{figure}[t]
\centering
 \resizebox{\hsize}{!}{\includegraphics{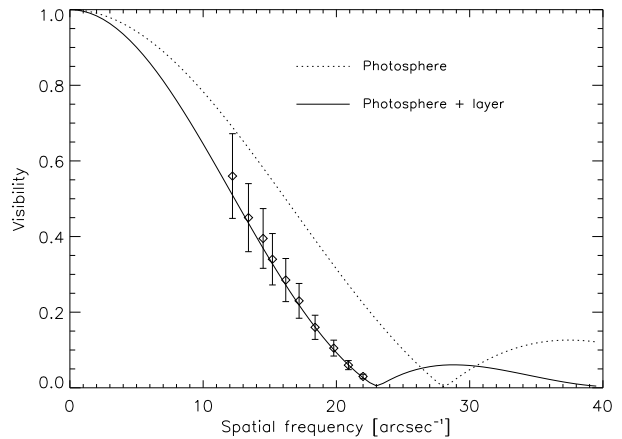}}
 \resizebox{\hsize}{!}{\includegraphics{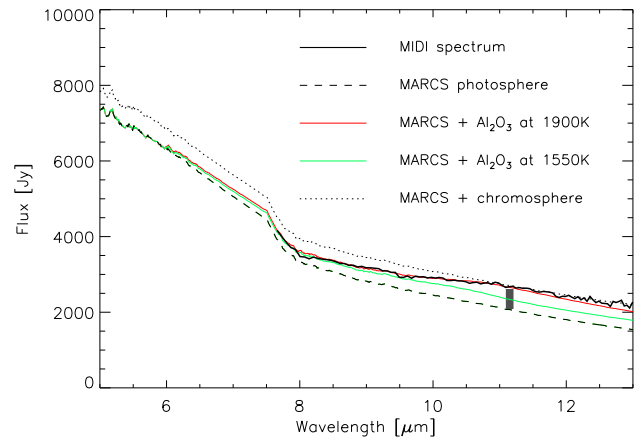}}
 \caption{
  By including $1 \times 10^{-3}$~g~cm$^{-2}$ of amorphous Al$_2$O$_3$
 at 1900\,K in the layer at 1.45~R$_{\star}$, we find a good fit to
 the ISI data (diamonds in the upper panel), and to the N-band excess
 (lower panel). The minor but systematic deviation at small spatial
 frequency suggests that the layer is actually located a little closer
 to the photosphere, i.e. 1.35~R$_{\star}$.  The grey box indicates
 the flux excess allowed by the flux ratio between the detached dust
 shell and the central object, i.e. the scaling factor used on the
 interferometric observations in the upper panel.  The excess of an
 optically thin isothermal chromosphere as in
 \cite{SkinnerWhitmore1987} is also shown (dotted line).
 }
\label{fig:ISI_al2o3}
\end{figure}

We have strong constraints on the spectral shape of the excess
emission and opacity: it should be transparent up to 8.5~$\mu$m, then
increase strongly up to about 12~$\mu$m and decrease again toward
longer wavelengths (Fig.~\ref{fig:midi_spectra} and
\ref{fig:olivine_fit}).
In the optically thin regime, the excess due to chromospheric emission
is constant throughout the N band because the free--free opacity is
proportional to $\lambda^2$ and the chromospheric temperature is high
enough for the source function to be in the Rayleigh--Jeans
limit. Such a constant excess is not compatible with the observed
increase of the excess with wavelength (see also
Fig.~\ref{fig:ISI_al2o3}). In the optically thick regime, the observed
shape of the excess requires (1) the temperature of the chromosphere to
decrease strongly along the line of sight and (2) the opacity to reach
unity within the N band. Because the total excess is far
below that of a homogeneous, hot and optically thick chromosphere with
a diameter of about 1.5~R$_{\star}$, the chromosphere must be very
clumpy.

\begin{figure}[t]
 \centering
 \resizebox{\hsize}{!}{\includegraphics{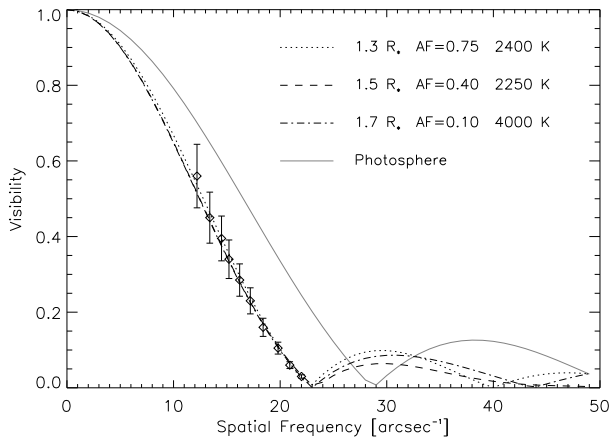}}
 \caption{
 Both the visibilities and the excess emission at 11~$\mu$m can be
 explained as due to a halo covered with hot spots. Allowed layer
 diameters range from about 1.3 to 1.8~R$_{\star}$. The derived hot
 spot temperatures remain well below the expected chromospheric
 temperature \citep[10000\,K,][]{Harper2001}. 
 }
\label{fig:chromosphere}
\end{figure}

To test this hypothesis, we constructed a model mimicking the central
photospheric disk embedded in a halo with optically-thick
chromospheric hot spots. This model yields the contrast ratio between
the mean\footnote{Note that the approximation of the spotted intensity
profile with a mean intensity profile yields reasonable visibility
curves only if the size of the spots is much smaller than the spatial
resolving power of the observations. This assumption is not violated
because (1) the observed visibilities are all within the first lobe,
and (2) they are all compatible with a single layer diameter and
therefore argue against large scale deviations from spherical symmetry
at these wavelengths.} observed photospheric intensity and mean layer
intensity as a function of layer diameter and chromospheric AF factor,
under the constraint that the chromospheric hot spots must generate
the excess emission seen at 11~$\mu$m in the MIDI spectrum and the
visibilities observed with the ISI. Fig.~\ref{fig:chromosphere} shows
how each assumed diameter of the chromospheric layer requires a
certain AF factor and hot spot temperature, if it is to match the
observed excess and visibilities. The AF factor goes down
asymptotically to zero for a layer radius approaching
1.8~R$_{\star}$. At the same time, the chromospheric temperature
increases to infinity, but it is clear that our model is too crude an
approximation of the actual intensity distribution on the sky in this
regime.

%
%
%

We find that for almost the entire range of possible layer diameters,
the hot spot temperature is far below the temperature of an actual hot
chromospheric component (10000~K). Only in the asymptotic regime
toward a layer radius of 1.8~R$_{\star}$ (80~mas on the sky) does the
model predict such a high temperature, at a very low AF factor (AF
$\le 0.03$), but the agreement with the observed visibilities is not
as good as with a smaller layer diameter. Moreover, it appears unlikely
that the very low AF factor can be combined with the requirements on
the source function.

We conclude that, while there must be some emission by a hot
chromospheric component in the mid IR, the {\sl hot spot temperatures
and required areal filling factor}, together with the {\sl spectral
shape of chromospheric emission}, appear to be in contradiction with
the hypothesis that an inhomogeneous hot chromosphere is responsible
for the observed N-band excess and diameter increase.

\paragraph{Dust}

The last viable candidate source for the extra opacity and emission is
dust.  The maximum of the excess emission appears to be located close
to 13~$\mu$m, which suggests a relation with the ``13~$\mu$m feature''
as seen in about 50 percent of all oxygen rich dust shells
\citep{Sloan1996} and attributed to a.o.  Spinel \citep[MgAlO$_4$,
][]{Posch1999}, SiO$_2$ \citep{Speck2000} or Alumina
\citep[Al$_2$O$_3$, ][]{Onaka1989}. In fact, all but the last
candidate have a fairly narrow, peaked absorption coefficient which
can not explain our observations.

%
%

A simultaneous modelling of the N-band excess and the ISI visibilities
with a layer of Al$_2$O$_3$ results in a very good agreement
(Fig.~\ref{fig:ISI_al2o3}) for a layer at 1900~K and a column density
of  $1 \times 10^{-3}$~g~cm$^{-2}$. Moreover, the transparency of
 Al$_2$O$_3$ in the near IR makes even the SiO band head at 7.7~$\mu$m well
visible through the layer, as required by the ISO-SWS spectrum. 

The derived temperature should be confronted with the temperature
regime only 0.5~R$_{\star}$ above the stellar photosphere. Although
the effective temperature of \aori is 3600~K, the outermost layers of
our {\sc marcs} model have temperatures of the order of only
2000~K. It appears therefore not unlikely that the region
0.5~R$_{\star}$ above the photosphere has a gas temperature of about
1900~K. However, demanding radiative equilibrium\footnote{We used the
dust radiative transfer code {\sc modust} and the optical constants of
amorphous Al$_2$O$_3$ measured by \cite{Begemann1997} and
\cite{Koike1995}}, i.e. the absorbed stellar radiation should be
emitted thermally, we arrive at a much higher temperature, of the
order of 2400~K.  

Al$_2$O$_3$ dust grains are believed to condense in chemical
equilibrium at a temperature of 1900~K only in high-pressure
environments, i.e. for a total pressure above $10^{-2}$~
bar \citep{Lodders1999}. This required pressure is a factor of $10^4$
larger than the pressure in the outermost layers of our {\sc marcs}
photosphere. 

\begin{figure}[t]
 \centering
 \resizebox{\hsize}{!}{\includegraphics{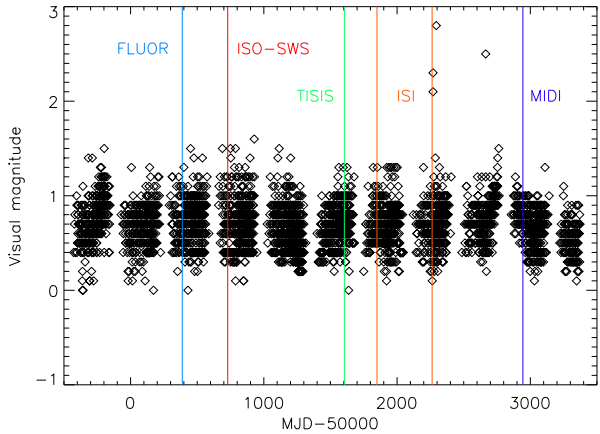}}
 \caption{
 Light curve of \aori at the time of the observations presented in this
 paper. Constructed from observations by the {\sl   American
 Association of Variable Star Observers (AAVSO)} (Waagen, E. O. 2004,
 private communication) 
 }
\label{fig:aori_lightcurve}
\end{figure}

\paragraph{Which one of the three?} A molecular origin can be ruled
out right-away as long as we assume that there have not been drastic
changes in the molecular environment of \aori between the epoch of the
ISO-SWS and the near-IR interferometric observations and that of the
mid-IR observations. The light curve of \aori
(Fig.~\ref{fig:aori_lightcurve}) shows no peculiar behaviour in that
time span. 

%
%
%

Chromospheric emission is certainly present, and we cannot rule out
some influence in the mid IR. However, as discussed above, we see no
way of reconciling the spectral and spatial constraints with a
chromospheric origin.

Dust appears to be the most attractive solution: it can reproduce
spectral and spatial observations simultaneously. Nevertheless, the
required temperature and pressure are a reason for concern. Furthermore, this
inevitably leads to questions on the relation between such an Alumina
layer and the detached olivine dust shell.

\subsection{Discussion}
\label{sect:disc_aori}
\begin{figure}[t]
\centering
 \resizebox{\hsize}{!}{\includegraphics{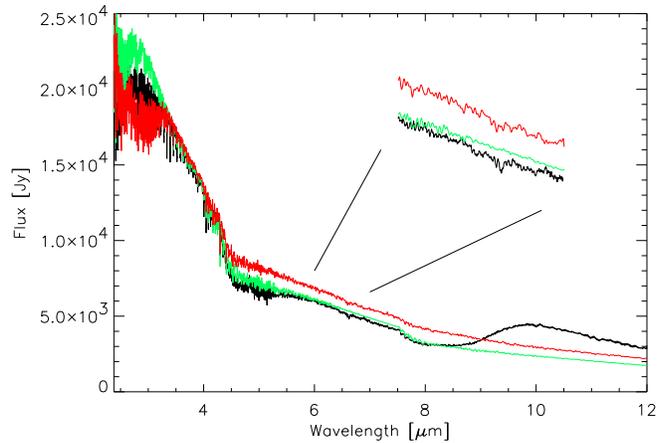}}
 \caption{
 ISO-SWS spectrum (black), {\sc marcs} photosphere model (green) and a
 photosphere+layer model (red) using the layer parameters by
 \cite{Ohnaka2004}. 
 }
\label{fig:ohnaka}
\end{figure}

\paragraph{The water layer}
Let us start this discussion by returning to the near-IR
observations. We found evidence for an optically thin molecular layer
containing \water and maybe also CO and OH about 0.45~R$_{\star}$
above the photosphere. The derived column density (about $2 \times
10^{19}$ cm$^{-2}$) is slightly higher than that derived by
\cite{Jennings1998}, but lower than that from \cite{Tsuji2000a} and
\cite{Ohnaka2004}, as listed in Table~\ref{table:aori_literature}.
The lower value found by \cite{Jennings1998} is the consequence of a
higher adopted water temperature: $2500-2800$~K. However, this
temperature is only an upper limit on the actual temperature of the
\water \citep{Jennings1998}.  \cite{Ohnaka2004} on the other hand
infers both a higher temperature and a significantly larger column
density. This is the consequence of his attempt to explain both the
mid-IR water spectrum at 6~and~11~$\mu$m and the ISI diameter at
11.15~$\mu$m as being due solely to a layer of \water.  
\begin{table*}
 \begin{center}
 \caption{
   Stellar and layer parameters for $\alpha$~Orionis, as derived by
   \cite{Jennings1998}, 
   \cite{Tsuji2000a}, \cite{Ohnaka2004} and \cite{RSG2004}. The
   first 2 determinations are purely spectroscopic, the third combines
   high-resolution spectroscopy and interferometry and the fourth one
   combines photometry with interferometry. (NA = not available)
   \label{table:aori_literature}
  }
 \vspace{3ex}
  \begin{footnotesize}
 \setlength{\tabcolsep}{.4mm}
 \begin{tabular}{lccccc}
  \hline
  \hline
  Parameter           & \cite{Jennings1998}        & \cite{Tsuji2000a}
  & \cite{Ohnaka2004} & \cite{RSG2004} & This work \\
  \hline
  $\theta_{\star}$ [mas] & NA     & NA                &  44.0
  & $42.00\pm 0.06$ & 45.6 \\
  T$_{\star}$ [K]         & NA    & 3600              & 3600
  & $3690 \pm 50$  & 3600   \\
  $\theta_{\rm{layer}}$ [mas] & NA  & NA                & $63.8 \pm
  6.4$    & $55.78\pm 0.04$ & 65  \\
  T$_{\rm{layer}}$ [K]     & $\le 2800$   & $1500 \pm 500$    & $2050 \pm 100$
  & $2055 \pm 25$ & 1750 \\
  Layer Composition    & H$_2$O  & H$_2$O            & H$_2$O
  & undefined   & H$_2$O, Al$_2$O$_3$(?)      \\
  K, opacity            & NA       & NA                & NA
  & $0.060\pm0.003$ & NA \\
  L opacity             & NA      & NA                & NA
  & $0.026\pm0.002$ & NA \\
  11$\mu$m opacity        & NA     & NA                & NA
  & $2.33\pm0.23$ & NA \\
  H$_2$O col. d. [cm$^{-2}$] &  $3\pm2 \times 10^{18}$ & $\sim 1 \times 10^{20}$ & $1-4
  \times 10^{20}$           & NA & $2 \times 10^{19}$ \\
 \hline
 \end{tabular}
\end{footnotesize}
 \end{center}
\end{table*}
Fig.~\ref{fig:ohnaka} compares a model using the layer parameters of
\cite{Ohnaka2004} with the ISO-SWS spectrum over the full wavelength
range. Improvements w.r.t. the model as presented by \cite{Ohnaka2004}
are the use of a {\sc marcs} photosphere instead of a blackbody
approximation and the use of the Ames line list for \water instead of
that from the {\sc hitemp} database. While this model reproduces quite
well the spectral features between 6 and 7~$\mu$m also used by
\cite{Ohnaka2004} (see inset of Fig.~\ref{fig:ohnaka}), it is not
compatible with the spectro-photometric flux levels. Furthermore, the
predicted band strength at 2.8~$\mu$m is too large and the SiO
band head at 8~$\mu$m, clearly seen in the ISO-SWS spectrum, is no
longer visible through such a thick layer.  We conclude that the model
as presented by \cite{Ohnaka2004} has too large an \water column
density.
%
%

\paragraph{The dust layer}

From the mid IR excess and the 11.15~$\mu$m interferometry, we found
evidence for the presence of hot amorphous Al$_2$O$_3$ at about the
same location above the photosphere as the \water layer. This is at
least a factor of 10 closer to the star than the onset of silicate
emission.

 Several dust condensation scenarios ascribe a crucial role to
Al$_2$O$_3$: it is assumed to be the first dust species to condense,
and thereafter act as a seed nucleus for further dust condensation
\citep[e.g.][]{Salpeter1977,Tielens1990,Sedlmayr1997,Lodders1999}.
Remark that \cite{Patzer2004} argue that this scenario is only valid in
chemical equilibrium conditions, and therefore not in the rapidly
expanding wind, but the conditions in the stationary layer we
consider here may be more favourable to attain chemical equilibrium.


If indeed {\sl we have observed the very onset of dust formation}, then it
is quite puzzling why no dust is seen in between 1.5 and
$10-20$~R$_{\star}$. If the mass loss is not episodic, there are only
two possibilities: either (1) there is no dust in this region, in
which case it must be destroyed right after its formation, or (2)
there is an outflow of amorphous Al$_2$O$_3$ but
at such a low density and temperature that we do not see it. 

 The former scenario is plausible given the presence of a patchy hot
chromospheric component which could destroy the Alumina before it
reaches cooler regions. The dust seen at larger distances may well be
formed anew, if the outflowing gas becomes again cool enough for
condensation to occur. The silicate emission then does come from a
detached shell.

The second hypothesis is supported by \cite{Onaka1989}, who propose
that the alumina grains are highly transparent up to the point where
they collect silicates on their surface. Coupled with a very low
density due to a strong acceleration at their birth in the
molecular layer, this would make them invisible up to the silicate
condensation location, where silicates settle onto the Al$_2$O$_3$
grains. We computed radiative equilibrium temperatures for both
silicates and  Al$_2$O$_3$ at the radius of the silicate condensation and
find the Alumina to be cooler by a few hundred K, making it indeed
undetectable\footnote{For the Alumina at the silicate condensation
radius to be detectable in the ISO-SWS spectrum, we would need a
Alumina mass loss rate which is 10 times higher than the silicate
mass loss.}! 

 This could mean that the dust shell of \aori is not really
``detached'' as previously assumed, but rather a continuous outflow
from close to the stellar photosphere, which is transparent up to the
point where silicates condense onto the alumina grains. Pure
Al$_2$O$_3$ is then only visible at fairly high temperatures near the
stellar surface where it is formed. 

 However, a major shortcoming in this scenario is that Alumina is
fairly transparent at short wavelengths and therefore radiation
pressure on a 0.01~$\mu$m grain is insufficient (by a factor of 10 at
least) to initiate the
outflow.


\section{Conclusions and prospects}
\label{sec:conclusions}

We have modelled the molecular (and possibly dusty)
close environment of the late-type supergiant $\alpha$~Orionis. We
took into account both spectral and spatial information from the near
to mid-IR. The improvements over previous modelling attempts for \aori are 
the use of a sophisticated {\sc marcs} model for the central star,
the computation of radiative transfer through the molecular layers in
spherical geometry, up-to-date line lists, and a larger set
of observational constraints. 

We find evidence for an optically thin layer of water close above the
photosphere. This layer gives rise to some spectral signature, but
does not increase the apparent size in the near-IR w.r.t. that of the
pure photosphere. However, in the mid-IR, we find excess emission by
amorphous silicates far out in the stellar wind (at least
20~R$_{\star}$ from the stellar surface) and another source of excess
emission much closer to the photosphere. The extra source of opacity
close to the star is so optically thick that it increases the apparent
size of the star with a factor of 1.5 from the near- to the mid-IR. It
must however be fully transparent up to 8~$\mu$m, and the excess
emission appears to decrease again beyond 15~$\mu$m. We show that it
cannot be of a molecular origin since that would induce strong
spectral features in the near IR. Chromospheric opacity/emission is
most definitely present at radio and UV wavelengths, but we see no way
to reconcile the spectral and spatial properties (inhomogeneity) of a
chromosphere with the near and mid-IR observations. Dust grains of
amorphous alumina (Al$_2$O$_3$) do yield a good spectral and spatial
fit, but at an uncomfortably high temperature (1900~K). Nevertheless,
this hypothesis fits in recent dust condensation scenarios and we
believe it to be the most likely solution.

New MIDI observations at 5 different baselines, high spectral
resolution and with simultaneous photometry are planned for autumn
2005 and, together with requested VISIR high spectral and spatial
resolution observations in the Q-band, will undoubtedly help to select
among the hypotheses.

\begin{acknowledgements}

The authors would like to thank the anonymous referee for his/her comments on
the presented dust condensation hypothesis. We also thank the people
from the ISI for making their data on $\alpha$ Orionis available.

\end{acknowledgements}

\bibliographystyle{aa}

\end{document}